# From EGEE Operations Portal towards EGI Operations Portal

Hélène Cordier, Cyril L'Orphelin, Sylvain Reynaud, Olivier Lequeux,
Sinikka Loikkanen and Pierre Veyre

IN2P3/CNRS Computing Center, Lyon, France

**Abstract** Grid operators in EGEE have been using a dedicated dashboard as their central operational tool, stable and scalable for the last 5 years despite continuous upgrade from specifications by users, monitoring tools or data providers. In EGEE-III, recent regionalisation of operations led the Operations Portal developers to conceive a standalone instance of this tool. We will see how the dashboard reorganization paved the way for the re-engineering of the portal itself. The outcome is an easily deployable package customized with relevant information sources and specific decentralized operational requirements. This package is composed of a generic and scalable data access mechanism, Lavoisier; a renowned php framework for configuration flexibility, Symfony and a MySQL database. VO life cycle and operational information, EGEE broadcast and Downtime notifications are next for the major reorganization until all other key features of the Operations Portal are migrated to the framework. Features specifications will be sketched at the same time to adapt to EGI requirements and to upgrade. Future work on feature regionalisation, on new advanced features or strategy planning will be tracked in EGI- Inspire through the Operations Tools Advisory Group, OTAG, where all users, customers and third parties of the Operations Portal are represented from January 2010.

## 1 INTRODUCTION

The need of a management and operations tool for EGEE and WLCG (Worldwide LCG) lead in 2004 to the creation of the EGEE Operations Portal, later referred to as "the CIC Portal"[1]. Its main focus is providing an entry point for all EGEE actors for their operational needs.

The portal is an integration platform as shown in fig.1, allowing for strong interaction among existing tools with similar scope and filling gaps wherever functionality was lacking. It also implements numerous work flows derived from procedures for several of its features out of requirements expressed by end-users or administrators of Virtual Organizations (VO), Regional Operations Centres (ROC) or Resources Centres.

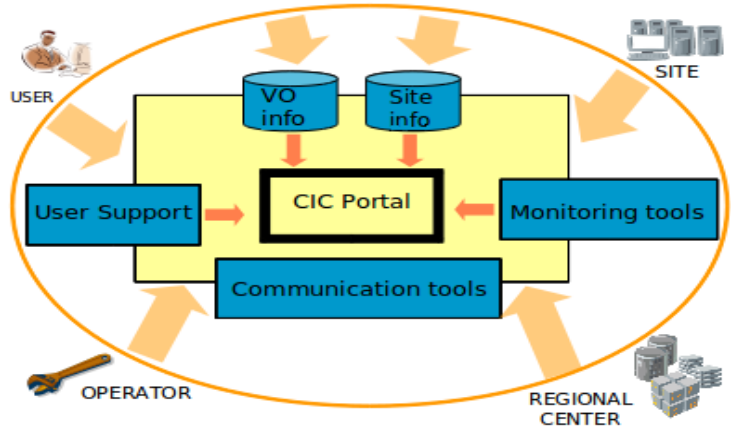

Fig.1: The integration platform concept

One of the most renowned representative features is the tool dedicated to daily operations of the grid resources and services through a set of synoptic views available to grid operators. After over 5 years of operations, this operations dashboard is now implementing a new decentralized operations model since June 2009, over a production infrastructure more than five times larger today than at the start of the project, i.e. over 250 sites and dozens of VOs.

## 2   EGEE OPERATIONS PORTAL: AN INTEGRATION PLATFORM

The Operations Portal is based on the "actor's view" principle where each EGEE actor has an access to information from an operational point of view according to his role in the project such as grid operator who daily monitors the status of resources and grid services, regular grid user and VO, site or ROC managers.

The information on display is retrieved from several distributed static and dynamic sources – databases, Grid Information System, Web Services, etc. – and gathered onto the portal. Linking this information enables us to display high level views where static and dynamic data yield representative views of the EGEE grid. This resulted in numerous tools that proved precious to sites like the User Tracking feature (to contact unknown users out of their DN in case of observed grid misuse) or the Alarm Notification feature (to subscribe to alerts upon monitoring failures).

Complementary to this informative goal, the portal also fosters communication between different actors, through channels like EGEE broadcast, and Downtime

Notification Mechanisms and putting in place procedures to address their interaction needs.

Finally, it offers an implementation of official operations procedures – see §3.1, activity reporting for sites and ROCs or VO life cycle creation and update for VO Management. Last but not least, some integration of third party tools resulting from tight collaboration, like monitoring probes re-submission tool SAMAP (SAM Admin Page) or sites configuration tool (Yaim Configurator) proved useful to operators and sites.

## 3   A SPECIFIC USE-CASE: THE OPERATIONS DASHBOARD

### 3.1   The COD story

When EGEE-I started, the infrastructure was smaller and managed centrally from the Operations Centre at CERN. While this worked quite well, troubleshooting of fifty sites was a difficult task, and the expertise of Grid Operations was concentrated in one place. In order to spread-out the expertise, several federations of countries shared shifts. Dubbed CIC-On-Duty (COD), this new system began in October 2004 [2], in which, the responsibility for managing the infrastructure was passing around the globe on a weekly basis. Indeed, this reduced the workload. However, requirements on tools synchronization and communication soared along with the complexity of the work. It then appeared necessary to have all the tools available through a single interface enabling a strongly interactive and integrated use of these tools. This resulted in the conception of a specific synoptic dashboard for operators and it became one of the main features of the Operations Portal dedicated to daily operations, the COD dashboard.

### 3.2   Operations dashboard concept

Accessing various tools from a single entry point is a key feature as the synoptic view and single operations platform proved invaluable gain for operators.

Indeed, the interface with the EGEE central ticketing system (Global Grid User Support, GGUS [3]), and the interface with its monitoring framework (Service Availability Monitoring, SAM [4] since December 2004 and Nagios [5] since December 2009), provide to the operators at the decentralized or at the central level a single entry point for:

- Detecting problems through the Operations dashboard, by browsing notifications triggered by the monitoring tool and checking administrative status of the site;

- Creating a ticket in EGEE ticketing system, GGUS [3], for a given problem, linking this ticket to the corresponding monitoring failure, and notifying the relevant recipients extracted from the Grid Operations Centre database (GOC database [6]) through customized template e-mails;
- Browsing, modifying, and escalating tickets from the Operations dashboard directly according to procedures;
- Communicating between teams and reporting to the Operations Coordination Centre.
- The main result at the end of EGEE-II was that the combined set of operational procedures and tools and their constant evolution have been recognized key in stabilizing sites in production.

### 3.3 *Overhaul of the dashboard feature during EGEE-III*

Since the early days of EGEE-III, the operational model evolved and the daily operations are now under the regions responsibility, even if still under the guidance of the operational procedures. The project still operates some supervision on the unattended problems at sites, urgent security matters or project wide operational issues.

**Status of the operations model in EGEE-III**
The COD dashboard has evolved accordingly over the last two years into an Operations Dashboard, enabling the daily operations at the federation level through "regional views" hosted centrally; the central layer indeed being reduced a minima. This reflects the evolution in the operational model of EGEE-III [7]. Tools and procedures are now adequate for the switch of the production infrastructure of EGEE-III, to a sustainable production environment based on the operational entities at the national level -National Grid Initiatives or NGIs.

However, the constraint to provide an interoperable back-end that can be distributed to existing federations, and at the national level in the next future, lead us to restructure the EGEE Operations Portal [8] at the same time on a global scale. With this reorganization during last summer, the Operations portal development team is now able to provide a customized and reliable tool, dealing with operations overview at national, federal or central level without any service disruption as exposed in fig.2.

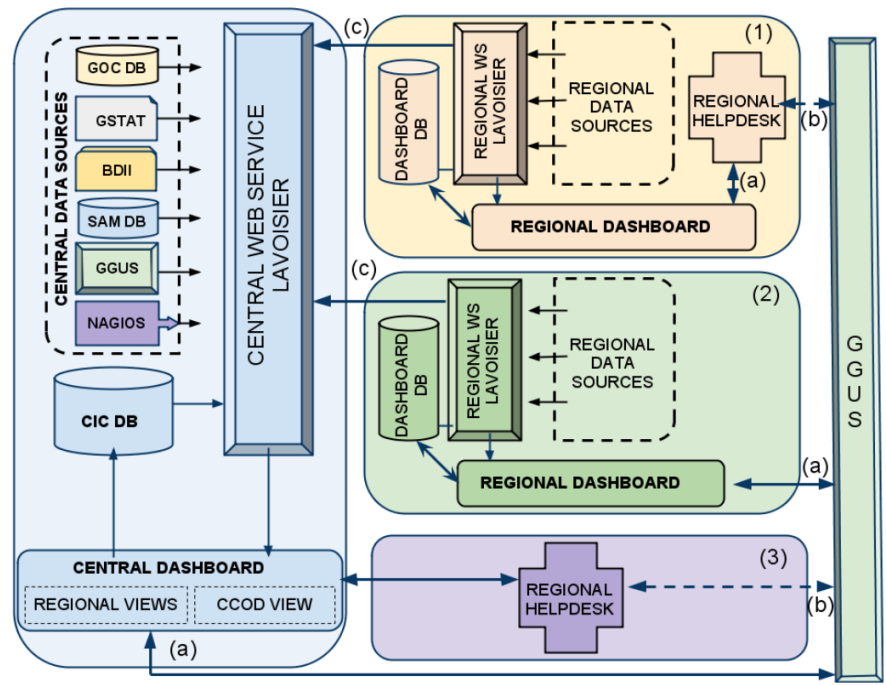

Fig.2. Operations dashboard concept
Legend of Fig.2:
(1) Full distributed model: regional help-desk, dashboard and database
(2) Partially distributed model : dashboard and database
(3) Help-desk distributed model
(a) update and creation of ticket
(b) synchronization between regional help-desk and GGUS
(c) synchronization between regional instance and central one

**Scalability, Adaptability, Reliability**
Together with the evolution of the operational model described above over the years, the portal also faced the evolution of all the third party tools it is coupled to, implying close interaction with their developers and administrators. The portal had been designed in such a way that neither the growth of the infrastructure to manage, nor the multiplication of the needs and procedures would overload or disrupt existing mechanisms. However, the range of tools proposed in the EGEE Operations Portal, becoming larger and larger, gradually induced an increased complexity in related development and maintenance.

To face this constant evolution and the EGI challenge on decentralisation and customisation as seen above, the global architecture of the portal evolved significantly and is summarized in §4.

**General Work-flow - Lavoisier**

We are highly dependent on the evolution of other tools: GOC DB, GGUS, SAM, Nagios (see fig. 2) and our technical solution was to implement a web service, in order to make the integration of these changes technically as transparent possible. Consequently, the main idea is to integrate each of the third party tools as a specific resource via Lavoisier.

"Lavoisier" [9, 10] has been indeed developed in order to reduce the complexity induced by the various technologies, protocols and data formats used by its data sources. It is an extensible service which provides a unified view of data collected out of multiple heterogeneous data sources. It enables to easily and efficiently execute cross data sources queries, independently of the technologies used. Data views are represented as XML documents [11] and the query language is XSL [12]. Lavoisier is distributed under the Apache 2.0 license like the Operations Portal.

The design of Lavoisier enables a clear separation in three roles: the client role, the service administrator role and the adapter developer role. The main client is the Operations Portal; it can retrieve the content of a data view in XML or JSON format through SOAP or REST requests. It can also submit XSL style sheets that will be processed by Lavoisier on the managed data views, and receive the result of this processing.

The service administrator is responsible for configuring each data view. He must configure the adapter, which will generate the XML data view from the legacy data source. He may also configure a data cache management policy, in order to optimize Lavoisier according to the characteristics and the usage profile of both the data source (e.g. amount of data transferred to build the view, update frequency, latency) and the generated data view (e.g. amount of data, time to live of the content, tolerable latency). Data cache management policy configuration includes:

- the cache type (in-memory DOM tree, on-disk XML file/files tree or no cache),
- view dependencies,
- a set of rules for triggering cache updates, depending on time-based events, notification events, data view read or write access, cache expiration, update of a cache dependency, etc.
- a set of fall-back rules for ignoring, raising errors or retrying cache updates in case of failure, depending on the type of the exception thrown,
- the cache time-to-live, to prevent from providing outdated information in case of successive cache update failures,
- a cache update time-out,
- the synchronization of the exposition of the new cache content for inter-dependant data views, in order to ensure data view consistency,

– the validation mode for generated views (conform to XML Schema [13], well-formed XML or no validation).

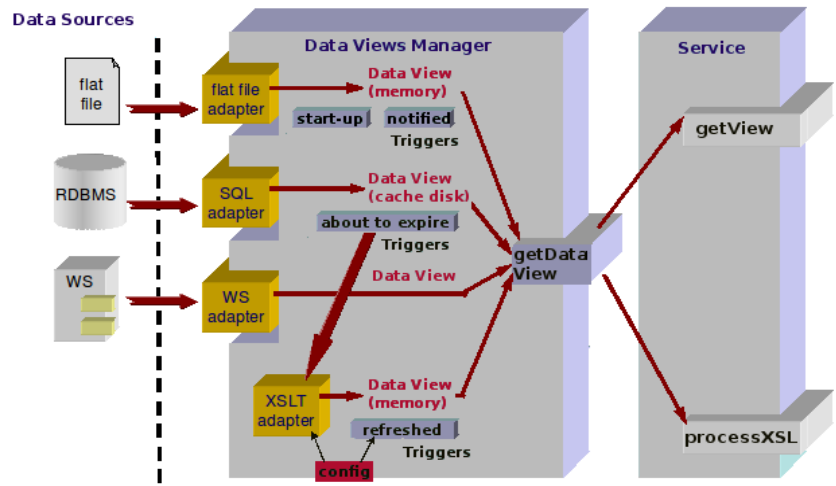

Fig. 3: Data Source Composition Service Lavoisier Configuration

Fig. 3 illustrates a very simple example of Lavoisier configuration with four data views. Data views obtained from flat file and RDBMS data sources are cached respectively in memory and on disk, while the data view obtained from the Web Service data source is regenerated each time it is accessed. The fourth data view is generated from the RDBMS data view, and refreshing of its in-memory cache is triggered when the cache of the RDBMS data view is refreshed. These two data views can be configured to expose their new cache simultaneously if consistency is required.

Lavoisier has proven effective in increasing maintainability of the Operations Portal, by making its code independent from the technologies used by the data sources and from the data cache management policy. Its design and interfaces made easier writing reusable code, and good performances are easily obtained by tuning the cache mechanisms in an absolutely transparent way for the portal code. Indeed, the different components work in a standardized way through the output of the Lavoisier Web Service. The translation of resource information into this standardized output is provided by different plug-ins.

Recently, a joint effort with EELA-II has been recently put into the configuration of Lavoisier, the structure of its caches and the rules of refreshing to have efficient, scalable and reliable data handling. Indeed, all information has been structured around the base component of the Operational Model: the site. We retrieve the global information about primary sources like GGUS, GOC DB, SAM and we organize it by sites. The main idea is to construct a summary of the different available information for a site: firstly, this organization permits to continue to work with the caches, even if a primary source is unavailable; then users access only information they need on the web page. Users' information is structured around a synoptic view of the site so they do not access the primary sources numerous times but a subset of them through the site view. Finally, we refresh the data sources only as needed and only when an action has been triggered. Last but not least, it is very easy to add a new data source in this model. In the configuration file of Lavoisier people can add the access to the relevant primary source and also the per site information split. This information is then readily available in the site synoptic view.

**Web portal architecture**
The advantages of adopting a framework are numerous. Namely, a framework streamlines application development by automating many of the patterns employed for a given purpose. It also adds structure to the code, prompting the developer to write better, more readable, and more maintainable code. Moreover, a framework makes programming easier, since it packages complex operations into simple statements. Additionally, several key features of a framework optimize the development of web applications, as these enable to:

- separate a web application's business rules, server logic, and presentation views,
- contain numerous tools and classes aimed at shortening the development time of a complex web application, and
- automate common tasks so that the developer can focus entirely on the specifics of its application.

On the other hand, adopting a framework induces a steep and long learning curve in itself. Also checking all the complex specifics - present and in the medium term - of operation portal was a pre-requisite before the start of the migration could take place. Proper feasibility and appropriate development of the operations dashboard started in August 2009 and was available for testing in December 2009

Indeed, the user interface is based on a View-Controller pattern and all the user requests are filtered out and checked by the main controller.

Whenever authorisation is needed, authentication is done using X509 certificate. Accepted requests are then forwarded to the sub-controller in charge of loading the requested page. In addition, the main controller uses abstraction layers to transparently connect to third-party applications like databases or web services. Connections are shared by all the sub-controllers thanks to inheritance mechanisms.

This approach improves organisation of source code, leading to a high rate of re usability. Moreover, all the operating system dependent implementations have been removed in order to ease configuration and deployment. It means also that the dashboard will be provided with a data schema and that with the help of the Symfony Framework you can generate on the fly the Database corresponding to the schema. This option is working with MySQL, Oracle, PostgresSQL and SQLLite.

The Symfony Framework is also giving additional features like a security layer assuming XSS and CSRF Protection, a set of different work environments different: test, prod and dev, and the use of plug-ins developed by the Symfony community. Finally, it has been thoroughly tested in various real-world projects, and is actually in use for high-demand e-business websites.

## 4   EGEE/EGI OPERATIONS PORTAL STATUS AT THE END OF EGEE-III

The portal architecture is still three-fold [8]: php web portal, database, and Data Processing System named Lavoisier [9,10]; however, the php web interface is based now on a PHP Framework, Symfony [14] and the Data Processing system Lavoisier has recently been enhanced to increase the speed in the information handling. Taking advantage of the php framework features, the code factorization and reusability has increased. Also, the framework structure has eased the build-up of the local packages of the dashboard for which information sources and data base need customization.

The operations dashboard is the first feature of the portal to undergo such a overhaul to prepare for potential decentralisation and subsequent customization. On the 9th of December 2009, the reorganization of the operations dashboard has been completed and the Nagios based operations dashboard has been available for testing; it had been validated and released on March 1rst 2010.

Decentralization for the time being is achieved via "views hosted centrally" and the relevant operational entities – Regional Operation Centers in EGEE and National Grid Initiatives in EGI will be able to run their own instance of the tool upon request at the end of March 2010. In the mean time a dedicated beta-package is available and had been already tested by a couple of federations [15].

### *4.1   Installation Requirements for Decentralized Package*

Any grid infrastructure will be able to install an Operations dashboard customized to its needs with a minimum set of requirements. A dedicated server will have to run PHP with a version over 5.2.4 with modules enabled such as SOAP and OCI, together with Java with a JDK version over 5.0 and a database of their choice MySQL, SQLLite, Oracle or PostgresSQL. The full package delivered will comprise the PHP code, the Database schema and the Lavoisier module, together with the relevant configuration files to proceed to necessary customization.

### *4.2 Conclusion and description of future work in EGI -Inspire*

Given our technical choices based on standards, we should be able to meet any scenario and associated requirements from any future operational entities like grid projects, federations or nations. Collaboration is already under way with EELA as a first example of adaptation to external DCI [16, 17, and 18].

In the mean time, current overhaul work comprises similar reorganization of the other features of the EGEE Operations Portal starting with EGEE broadcast and VO life cycle and VO operational metrics. We will then proceed with other key features.

In parallel, our goal is to provide a seamless access to the users regarding the sites' and the VOs' info, an easy access to administration interfaces for sites and VO managers, a standard access to information and standard formats for it, together with a scalable operations model for each scenario potentially required: national, federal or central.

Consequently, for user ergonomics and back-end simplicity the operations Portal development team is already working on common development plans with GOCDB developers [19]. Both developers' teams are part of the Operations Tools Advisory Group, OTAG, in order to cope with EGI/NGIs new features requirements and priorities. This newly formed advisory group will be transversal to all operational tools in EGI. Namely, potential decentralization of operational tools, migration and re-engineering of the existing features of the operations portal recognized as key for EGEE/EGI operations as well as strategic decisions are done in agreement with this joint Advisory Group: OTAG [20].

Finally, the structure of the back-end of the Operations Portal enables now any operational entity (NGIs or group of NGIs) to potentially customize their own configuration with the database of their choice, the relevant Lavoisier plug-ins and their specific set of php files. It is worth quoting that Lavoisier plug-ins developed externally -when developed in a generic way- might be reusable and made available to the community. Global trend clearly indicates customization as the motto for us to claim: we are aiming at delivering flexible services and customizable end-user interfaces.

## 5   ACKNOWLEDGEMENT

This work makes use of results produced with the EGEE (http://www.eu-egee.org) grid infrastructure, co-funded by the European Commission (INFSO-RI-222667)